\documentclass[useAMS,usenatbib]{mn2e}

\usepackage{txfonts} 
\usepackage{epsfig} 
\usepackage{graphicx}

\newcommand{\dechms}[4]{$#1^{\rm h}#2^{\rm m}#3\mbox{$^{\rm s}\mskip-7.6mu.\,$}#4$}

\newcommand{\decdms}[4]{$+#1^{\circ}#2'#3\mbox{$''\mskip-7.6mu.\,$}#4$}

\title[A spider-like outflow in Barnard 5 - IRS 1]{A spider-like outflow in Barnard 5 - IRS 1:
 The transition from a collimated jet to a wide-angle outflow?}
\author[Zapata et al.]{Luis A. Zapata$^{1}$\thanks{E-mail: lzapata@crya.unam.mx}, 
H\'ector G.~Arce$^{2}$,  Erin Brassfield$^{3}$, Aina Palau$^4$, 
\newauthor
Nimesh Patel$^{3}$, and Jaime E. Pineda$^5$\\ \\
$^{1}$Centro de Radioastronom\'\i a y Astrofis\'\i ca, UNAM campus Morelia, M\'exico \\
$^{2}$Department of Astronomy, Yale University, P.O. Box 208101, New Haven, CT 06511, USA \\
$^{3}$Harvard-Smithsonian Center for Astrophysics, 60 Garden Street, Cambridge MA 02138, USA\\
$^{4}$Institut de Ci\`encies de l'Espai (CSIC-IEEC), Campus UAB-Facultat de Ci\`encies, Torre C5-parell 2, 
E-08193 Bellaterra, Catalunya, Spain\\
$^{5}$ Institute for Astronomy, ETH Zurich, Wolfgang-Pauli-Strasse 27, CH-8093 Zurich, Switzerland}
\begin{document}

\date{Sent 2013}

\pagerange{\pageref{firstpage}--\pageref{lastpage}} \pubyear{2013}

\maketitle

\label{firstpage}

\begin{abstract}  
We present line and continuum observations made with the Submillimeter Array (SMA) of the
young stellar object Barnard 5 - IRS1 located in the Perseus molecular cloud. Our $^{12}$CO(2-1) 
line observations resolve the high-velocity bipolar northeast-southwest outflow  
associated with this source. We find that the outflowing  gas shows different structures at three 
different velocity regimes, in both lobes, resulting in a spider-like morphology.
In addition to the low-velocity, cone-like (wide-angle) lobes that have previously been observed, 
 we report the presence of intermediate-velocity parabolic  shells emerging very close 
to the Class I protostar, as well as high velocity molecular bullets that appear to be associated to the  
optical/IR jet emanating from this source.
These compact high-velocity features reach radial velocities of about 50 km s$^{-1}$ away from the cloud velocity.
We interpret the peculiar spider-like morphology is a result of the molecular 
material being entrained by a wind with both a collimated jet-like component 
and a wide-angle component.  
We suggest the outflow is in a transitional evolutionary phase 
between a mostly jet-driven flow and 
an outflow in which the entrainment is dominated by  the wide-angle wind component.
%in which the velocity of the outflowing material depends on the angle from the outflow axis. 
%Furthermore, we suggest this morphology is produced by an episodic outflow 
%that is in a transitional evolutionary phase 
%between a mostly jet-driven flow and an outflow in which the entrainment is dominated by  the wide-angle wind component.
We also detect 1300 $\micron$  continuum emission at the position of the protostar, which 
likely arises  from the dusty envelope and disk surrounding the protostar.
Finally, we report the detection of $^{13}$CO(2-1) and SO(6$_5$-5$_4$) emission arising from the 
outflow and the location of the young stellar object.
\end{abstract}


\begin{figure*}
\begin{center}\bigskip
\includegraphics[scale=0.16]{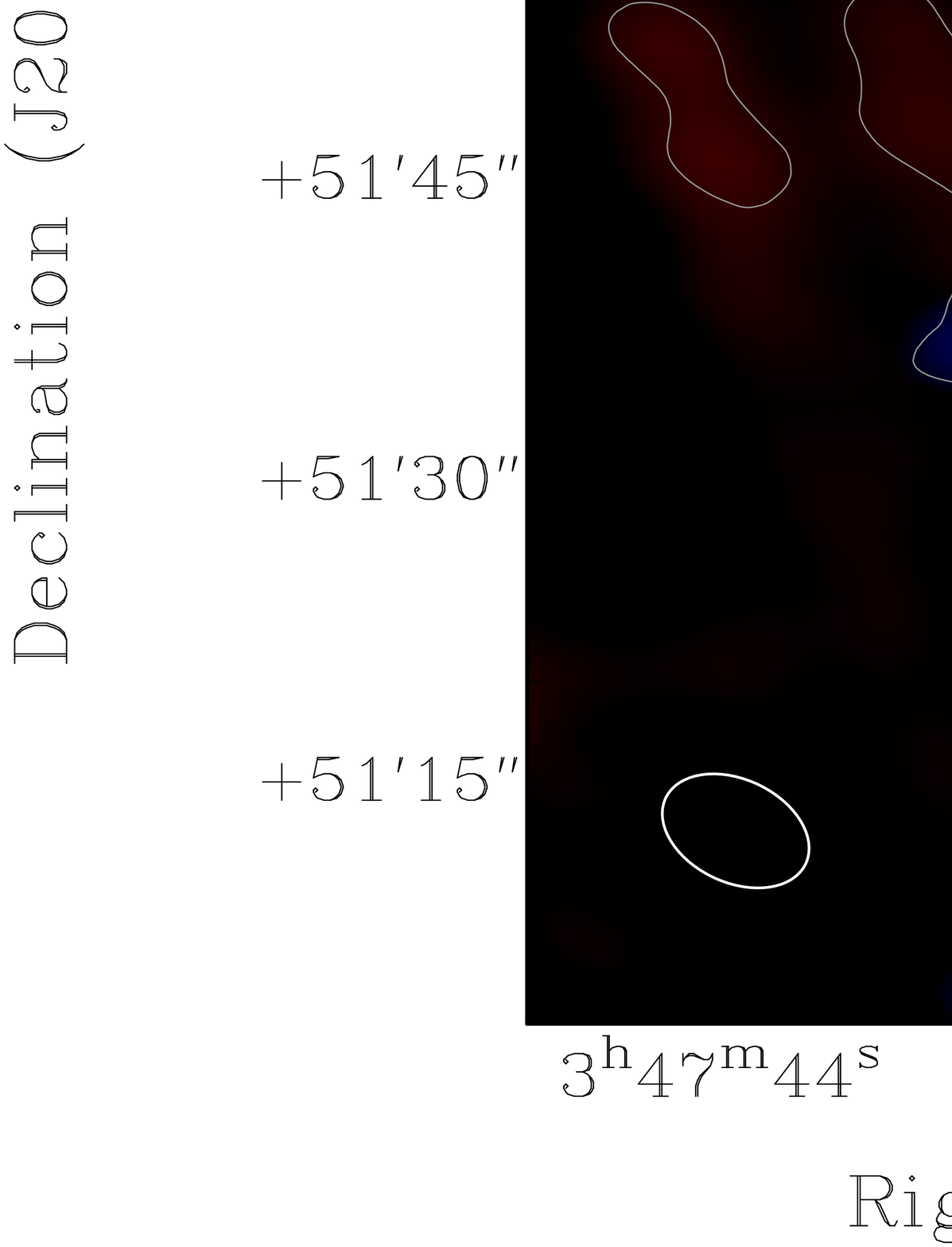}
\includegraphics[scale=0.18]{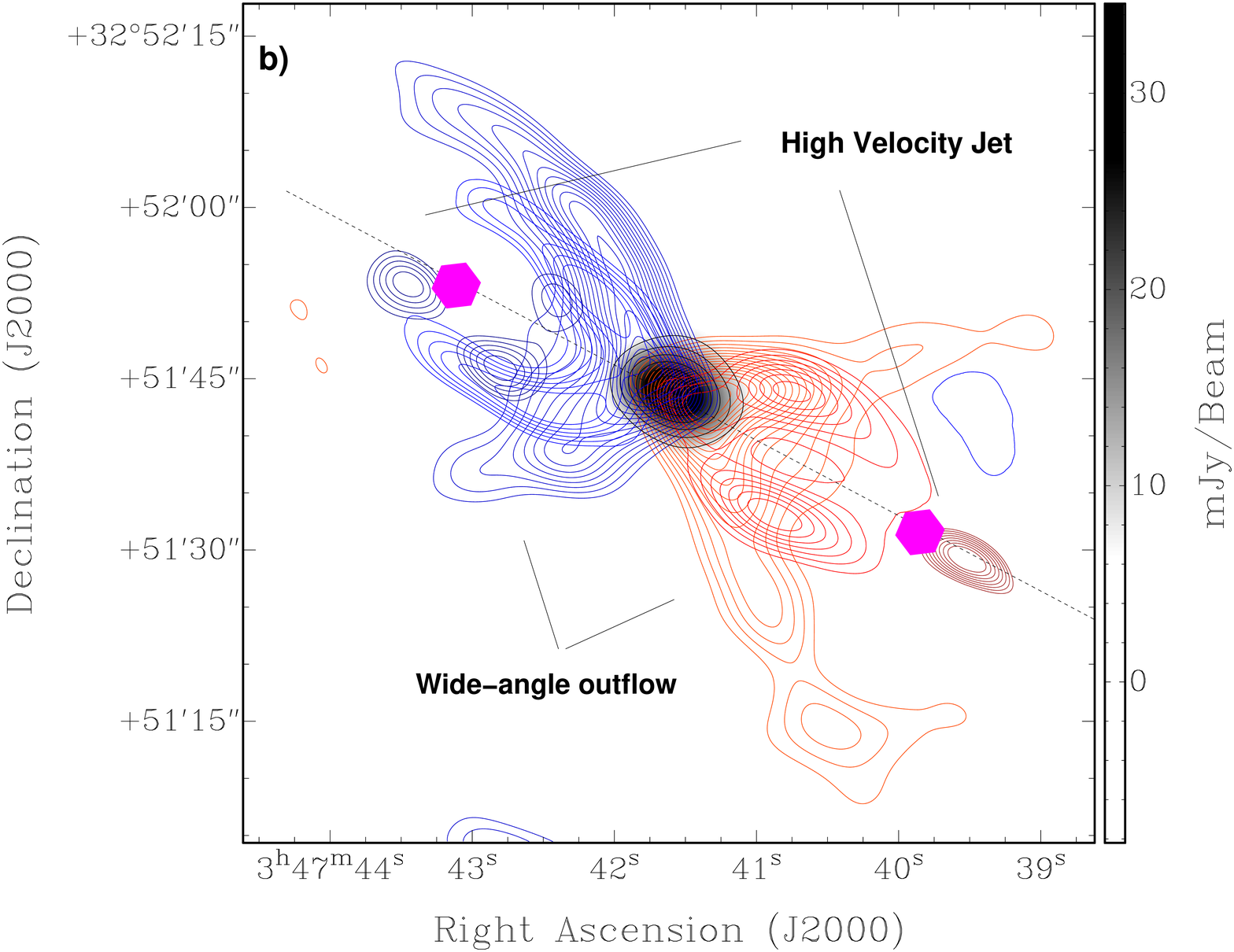}
\caption{
                {\bf a)} SMA integrated intensity (moment 0) color scale and contour map of the $^{12}$CO(2-1) 
                 emission (grey contours), and the continuum emission at 1300 $\micron$ (white contours) 
                from B5-IRS 1.  The blue color scale represents blueshifted gas, while
                the red color scale represents redshifted gas. The integrated velocity range for the blueshifted side is from $-$50.9 to  $+$6.2 km s$^{-1}$,
                while for the redshifted side is from $+$14.7 to $+$70.8 km s$^{-1}$.
                The white contours  
                range from 35\% to 85\% of the peak emission, 
                 in steps of 10\%. The peak of the continuum emission is 0.032 Jy beam$^{-1}$. 
                 The grey contours range from 10\% to 87\% of the peak emission, 
                 in steps of 7\%. The peak of the $^{12}$CO(2-1) line emission is 150 Jy beam$^{-1}$ km s$^{-1}$.
                 The emission peak here is the arithmetic sum of the total integrated CO emission. 
                 The synthesized beam of the continuum image is shown in the lower left corner.               
                 {\bf b)} Integrated intensity map of the Barnard 5 IRS1 $^{12}$CO(2-1) outflow 
                 at different velocity ranges.
                The different blue and red color tones of the contours represent different range of velocities 
                 with the lighter (darker) hues representing lower (higher) outflow velocities. 
                 The low-velocity blue and red ranges are  [$-$4.3 to $+$6.2  km s$^{-1}$] and [$+$14.7 to $+$24.2 km s$^{-1}$]. 
                 The intermediate-velocity ranges are [$-$39.0 to $-$15.9 km s$^{-1}$] and [$+$35.1 to $+$59.7  km s$^{-1}$], 
                 while the high-velocity ranges are [$-$50.9 to $-$40.3  km s$^{-1}$] and [$+$60.2 to $+$70.8 km s$^{-1}$]. 
                 The high velocity bullets show a larger spread of velocities (see Figure 4), however,
                for clarity, we choose these slightly narrower high-velocity windows.  
                The contours for the different blue and red contours 
                 range from about 30\% to 90\% of the peak emission, in steps of 5\%. 
                 In the high-velocity emission the contours start from 50\% the peak 
                 of the line emission. 
                 The two pink hexagons mark
                 the positions of the strongest H$_2$ and H$\alpha$ knots reported by \citet{yu1999}.
              }
\label{fig2}
\end{center}
\end{figure*}

\section{Introduction}

Bipolar outflows are an ubiquitous phenomena associated with star formation.
 The leading theories on the origin of protostellar outflows indicate that they are generated through
the interaction of ionized disk material and the  magnetic fields from the forming star and/or disk
  \citep[and references therein]{li2014}.
Protostellar winds entrain the host cloud's molecular gas thereby producing molecular outflows \citep[e.g.][]{arce2007}.

There is substantial observational evidence showing  
that molecular outflows from low-mass young stars tend to have relatively well-collimated lobes early 
in their development and then evolve to have wider opening angle flows at later stages \citep{rich2000,arce2006,sea2008}. 
However, there is no consensus on the detailed physics that produce this evolutionary trend in molecular outflows.
It has been proposed that  jet axis wandering (precession) could produce wider cavities as the protostar 
evolves \citep{mass1993}. However, in most sources the opening angle of the precession cone is smaller 
than the observed wide-angle outflow cavity \citep[e.g.,][]{rei2000}. A second possibility is that protostellar 
winds have both a collimated and a wide-angle component, and the observed molecular outflow is 
predominantly driven by one of the two components, depending on the age of the protostar 
\citep[e.g.,][]{sha2006,rom2009}. In this scenario the  
opening angle and axis direction of the wind arising from the
protostar/disk system remain approximately constant, but the wind increasingly 
disperses more of the circumstellar gas, thereby creating wider outflow cavities, as the protostar evolves and 
the circumstellar envelope decreases in density \citep[e.g.,][]{off2011}.

Barnard 5 - IRS 1 (hereafter, B5-IRS1) is a young stellar object located at the eastern end of the Perseus molecular cloud complex 
at a distance of about 240 pc  \citep{hiro2011}. This object is embedded in a dense core with a filamentary structure \citep{pin2011}. 
B5-IRS1 has a bolometric luminosity of $\sim$ 5 L$_\odot$ \citep{bei1984,eva2009} 
and drives a giant bipolar flow associated with two clusters 
of HH objects: HH 366 E to the northeast and HH 366 W to the southwest \citep{bally1996, yu1999}. 
A molecular outflow with a bipolar wide-angle morphology and  a southwest-northeast orientation
is found at the base of these HH objects \citep{full1991,lan1996,vel1998}.
Both the blueshifted lobe (northeast of the source) and the redshifted lobe (southwest of the source) 
show a wide-angle cone-like structure with a projected opening angle of about 90$^\circ$ \citep{lan1996}. 
Near infrared and optical images shown by  \citet{yu1999} reveal
H$_2$ and H$\alpha$ emission within 20$''$ of  B5-IRS1,
arising from a jet that bisects the limb-brightened CO cones.
\citet{yu1999} suggested that the presence of both axial jet-knots and a wide-angle cavity implies that the 
central source may simultaneously power both a jet and a wide-angle wind.

In this paper, we present new line and continuum Submillimeter Array\footnote{The Submillimeter Array (SMA) 
is a joint project between the Smithsonian Astrophysical Observatory and the Academia Sinica Institute of Astronomy and 
Astrophysics, and is funded by the Smithsonian Institution and the Academia Sinica.} (SMA) observations of the young 
star B5-IRS1. These observations reveal further structure in the bipolar outflow associated with this young stellar object 
and show that it has a peculiar morphology.

\section[]{Observations}

The observations were made with the SMA, and were collected on 2007 September 13, 
when the array was in its subcompact configuration.  The 21 independent baselines in this
configuration ranged in projected length from 5 to 35 k$\lambda$.  The phase reference center 
for the observations was set to $\alpha_{J2000.0}$ = \dechms{03}{47}{41}{60}, $\delta_{J2000.0}$ = 
\decdms{32}{51}{43}{5}. Two frequency bands, centered at 230.53799 GHz (Upper Sideband) and 
220.53799 GHz (Lower Sideband) were observed simultaneously. The primary beam of the SMA at 
230 GHz has a FWHM of about $50''$, and the continuum emission arising from B5-IRS1
falls well within it. The line emission, on the other hand,  extends beyond the FWHM of the primary beam, 
so we corrected the images for the primary beam attenuation. 

The SMA digital correlator was configured to have 24 spectral windows (``chunks'') of 104 MHz and
128 channels each. This provided a spectral resolution of 0.812 MHz ($\sim$ 1 km s$^{-1}$) 
per channel.  Observations of  Uranus provided the absolute scale for the flux density calibration.  
The gain calibrators were the quasars 3C 111 and 3C 84, while 3C 454.3 was used for bandpass calibration. 
The uncertainty in the flux scale is estimated to be between 15 and 20$\%$, based on the SMA monitoring of 
quasars. 

The data were calibrated using the IDL superset MIR, originally developed for the Owens Valley 
Radio Observatory \citep[OVRO,][]{Scovilleetal1993} and adapted for the SMA.\footnote{The 
MIR-IDL cookbook by C. Qi can be found at http://cfa-www.harvard.edu/$\sim$cqi/mircook.html.} 
The calibrated data were imaged and analyzed in the standard manner using the {\emph MIRIAD} \citep{sau1995} and
{\emph KARMA} \citep{goo96} softwares\footnote{The calibrated data can be obtained from: 
http://www.cfa.harvard.edu/rtdc/sciImages/070913\_125512\_b5irs1.html}.  
A 1300 $\micron$ continuum image was obtained by averaging line-free 
channels in the lower sideband with a bandwidth of about 2 GHz. 
For the line emission, the continuum was also removed.
For the continuum emission, we set the {\emph ROBUST} parameter of the task {\emph INVERT} to 0 to obtain 
an optimal compromise between resolution and sensitivity, 
while for the line emission we set this to -2 in order to obtain a higher angular
resolution. The resulting r.m.s.\ noise for the 
continuum image was about  2 mJy beam$^{-1}$ at an angular resolution of $7\rlap.{''}74$ 
$\times$ $4\rlap.{''}87$ with a P.A. = $66.3^\circ$. The r.m.s.\ noise in each channel of the 
spectral line data was about 80 mJy beam$^{-1}$ at an angular resolution of 
 $7\rlap.{''}51$ $\times$ $4\rlap.{''}18$ with a P.A. = $63.2^\circ$.

\begin{figure}
\begin{center}\bigskip
\includegraphics[scale=0.4]{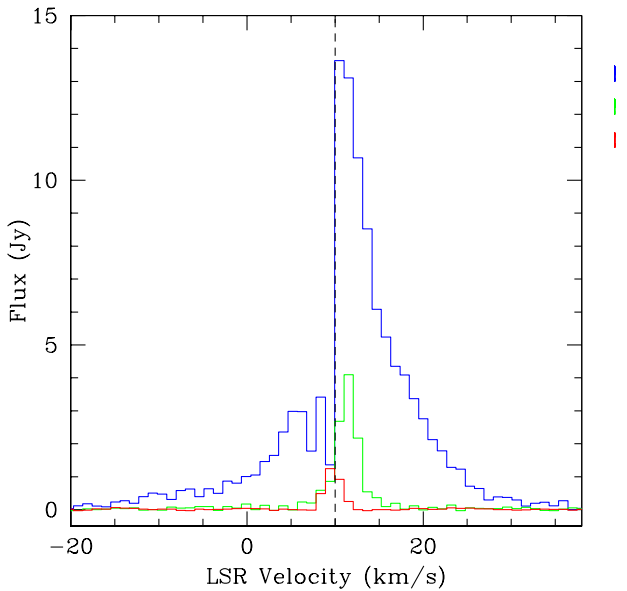}
\caption{
               Sum spectra of the three lines detected towards B5-IRS1:
                  $^{12}$CO(2-1) (blue), $^{13}$CO(2-1) (green), and
                  SO(6$_5$-5$_4$) (red).
                The sum spectra were obtained from an area delimited by a
               $45"\times45"$ square for $^{12}$CO(2-1), and  a
               $20"\times20"$ square for $^{13}$CO(2-1) 
               and SO(6$_5$-5$_4$), all centered at the position of the source. 
               The vertical dashed line represents the location of the
               systemic velocity of the cloud, V$_{LSR}$ $\sim$ 10.0 km s$^{-1}$.   
               The spectra were obtained with 
               the task {\emph VELPLOT} of {\emph MIRIAD}.        
              }
\label{fig1}
\end{center}
\end{figure}

\section[]{Results and Discussion}

\subsection[]{Continuum}

In Figure 1, we show the 1300 $\micron$ continuum image resulting from these millimeter SMA observations. 
We only detect a single source that is associated with B5-IRS 1 at the position 
of $\alpha_{J2000.0}$ = \dechms{03}{47}{41}{5}, $\delta_{J2000.0}$ = 
\decdms{32}{51}{43}{8}, with a positional error of less than 1\rlap.{"}0.
From a Gaussian fit to the continuum emission we obtain that 
the flux density and peak intensity values of this compact source at this wavelength 
are 54$\pm$8 mJy  and 31$\pm$5 mJy beam$^{-1}$, respectively.  
The Gaussian fit was also used to determine that the deconvolved size for this continuum source is 
$5\rlap.{''}5$ $\pm$ $0\rlap.{''}5$ $\times$ $4\rlap.{''}0$ $\pm$ $0\rlap.{''}6$  with a P.A. = $-$28$^\circ$ $\pm$ 10$^\circ$.
Therefore, at the distance of the Perseus molecular cloud complex
the size of the continuum source is about 1200 AU.

Following \citet{hil1983} and assuming optically thin isothermal dust emission, a gas-to-dust ratio of 
100, a dust temperature of 30 K, a dust mass opacity $\kappa_{1300 \micron}$ 
= 1.1 cm$^2$ g$^{-1}$ \citep{oss1994},  %an emissivity index $\beta = 1.0$ \citep{lan1996}, 
and that this object is located at 240 pc, we estimate  
the total mass associated with the dust continuum emission to be 0.02 M$_\odot$, this in a good agreement 
with the value obtained by \citet{bra2011}. However, we remark that the mass estimate presented in \citet{bra2011}
was obtained using a restricted UV-range trying to avoid the envelope contribution. 

The millimeter continuum emission is probably tracing the envelope and the circumstellar disk surrounding B5-IRS 1. 
Furthermore, we remark that the detected continuum emission could be partially optically thick, and therefore the reported mass estimate is a lower limit. 

\begin{figure}
\begin{center}\bigskip
\includegraphics[scale=0.4]{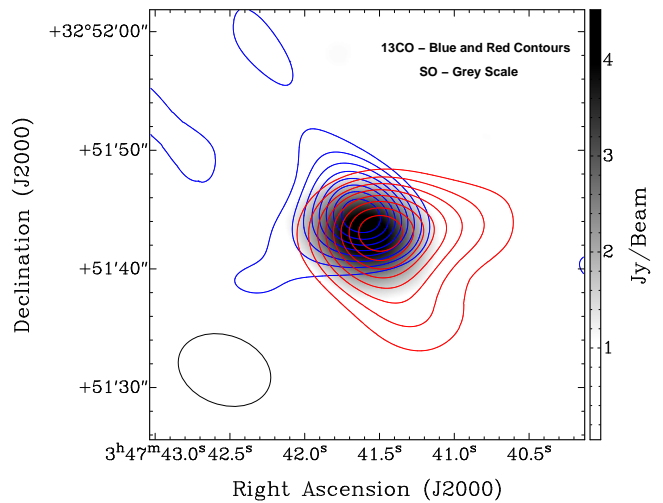}
\caption{ 
               Moment zero map of the SO line emission (grey scale) shown in combination with the blue- and red-shifted $^{13}$CO 
               emission (blue and red contours, respectively).
               For  SO(6$_5$-5$_4$) the velocity integration range is from $+$7 to $+$12 km s$^{-1}$. 
               For $^{13}$CO(2-1) the velocity integration range for the blueshifted lobe is from $+$6.5 to $+$8.6
               km s$^{-1}$, while for the redshifted lobe is from $+$11.8 to $+$16.0 km s$^{-1}$. 
               The blue and red contours range from 20\% to 90\% 
               of their respective peak emission,  in steps of 10\%. The peak of the blue and red emission are
               6.4 and 2.5  Jy beam$^{-1}$ km s$^{-1}$, respectively. 
               The synthesized beam of the SO and the $^{13}$CO lines is shown in the lower 
               left corner. The grey-scale bar on the right pertain to the SO line emission map.}
\label{fig4}
\end{center}
\end{figure}

\subsection[]{Spectral Lines}

In Figure 2, we show the lines that were detected in the two sidebands and that are discussed throughout 
this article. We simultaneously detected three lines, the $^{12}$CO(2-1) at a rest frequency of about 230.5 GHz, $^{13}$CO(2-1) 
at a rest frequency of about 220.5 GHz, and the SO(6$_5$-5$_4$) line at a rest frequency of about 219.9 GHz. The exact values
for the rest frequency of these lines can be found in the database for astronomical spectroscopy: {\emph http://splatalogue.net}.
The sum spectra of $^{13}$CO(2-1) and SO(6$_5$-5$_4$)  are narrower and fainter  compared to the $^{12}$CO(2-1) spectrum. 
The former have a  FWHM between 3-5 km s$^{-1}$, while the latter has a FWHM of 20 km s$^{-1}$.
The sum spectra of  $^{12}$CO and $^{13}$CO show  wings at high velocities that are attributed 
to the presence of molecular outflows. This is confirmed by our integrated intensity maps showing a clear bipolar 
structure, see Figures 1 and 3. 

\begin{figure*}
\begin{center}\bigskip
\includegraphics[scale=0.35]{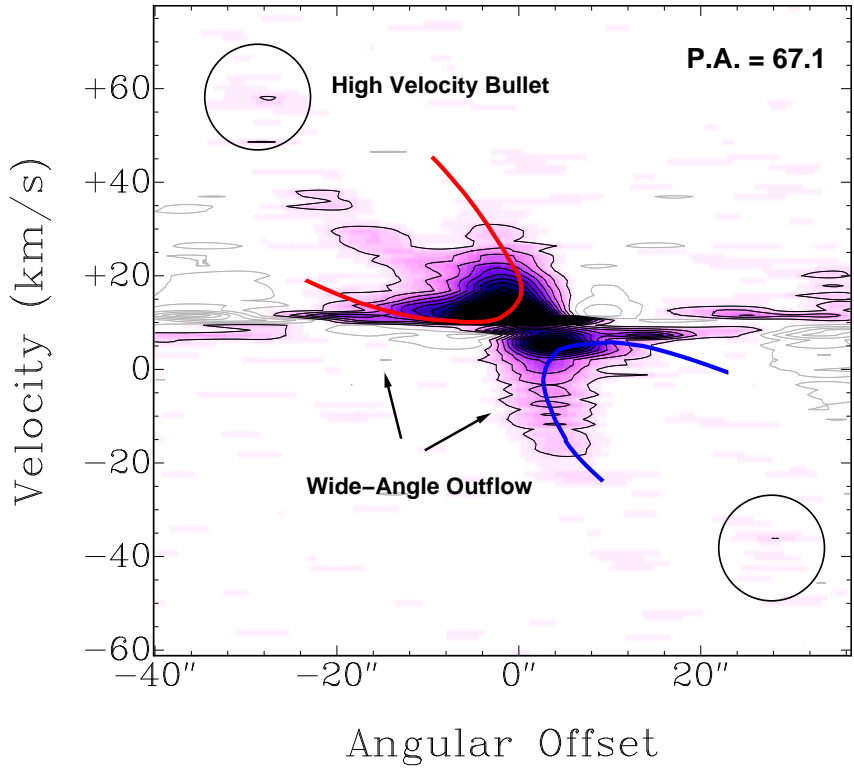}
\caption{
                Position velocity diagram (PVD) along the axis of the outflow that is shown in Figure 1.
                The PVD was made at a position angle of 67.1$^\circ$.  The black contours range from 6\% to 90\% 
                of the peak emission,  in steps of 4\%. The peak of the line emission is 6 Jy beam$^{-1}$. 
                The grey contours represent the negative
                emission arising mainly from velocities close to the cloud's systemic velocity 
                and have the same value as the black contours, but negative. 
               The curves mark the different structures found in our PVD. The circles mark 
                the highest velocity features in the outflow. Velocity is shown with respect to the Local Standard of Rest (LSR).  }
\label{fig3}
\end{center}
\end{figure*}

\begin{figure*}
\begin{center}\bigskip
\includegraphics[scale=0.4]{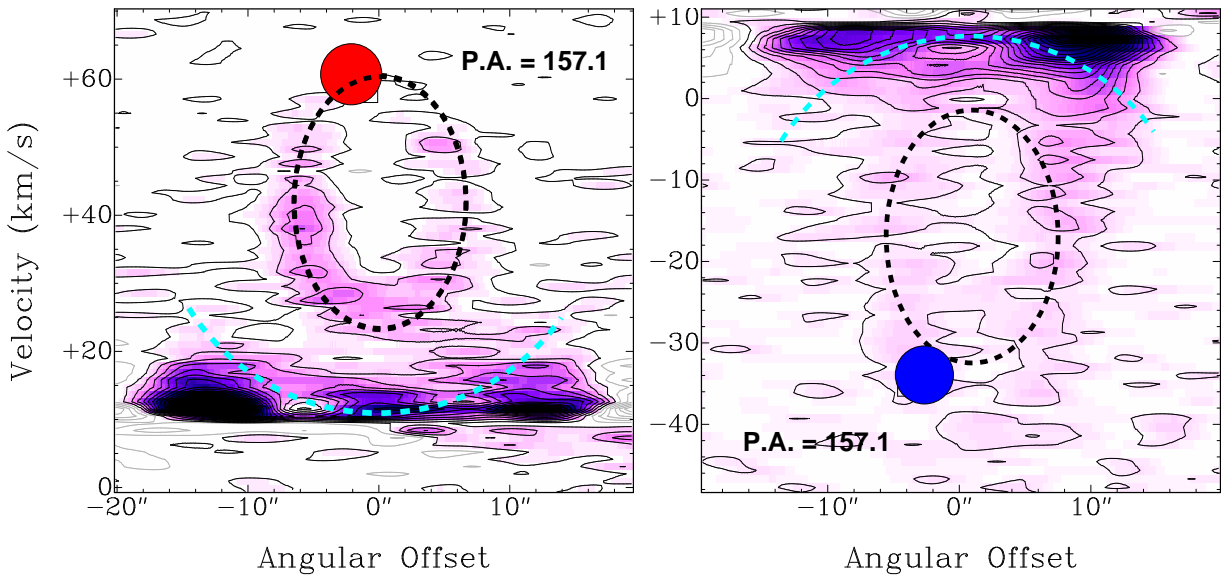}
\caption{
                Left panel: PVD perpendicular to the outflow axis in the redshifted lobe. The PVD was made at a position angle of 157.1$^\circ$.
                The contour levels and the peak emission are the same as Figure 4.
                Right panel: The same as in the left panel, but for the blueshifted lobe. 
                The contour levels and the peak emission are the same as Figure 4. 
                The blue and red dots represent the position and velocities of the high-velocity bullets shown in Figure 1{\it b}. The dashed curves mark 
                the different structures found in the PVDs. Both PVDs were made approximately 15$''$ away from the continuum source in both sides of the outflow.}
\label{fig3}
\end{center}
\end{figure*}

In Figure 1a we show the velocity-integrated intensity, or zero-moment map,  
of $^{12}$CO(2-1) for both the blue and red outflow lobes. 
The velocity range of integration for the blueshifted lobe is from $-$50.9 to $+$6.2 km s$^{-1}$, 
while for the redshifted lobe is from $+$14.7 to $+$70.8 km s$^{-1}$.
With our SMA observations we detect a substantially larger spread of velocities 
because the SMA data have significantly better sensitivity than previous studies  \citep[see ][]{vel1998}. 
The left panel in Figure 1 shows the wide-angle bipolar outflow that has been reported
 in previous high-angular resolution studies of this source \citep{lan1996,vel1998}. 
 In addition, our higher sensitivity and fidelity maps reveal new features in the outflow, 
 especially at high velocities, that are shown in Figure 1b. 

In Figure 1b, we split  the range of integration into three different velocity ranges 
 in both the blueshifted and redshifted lobes. The velocity ranges were selected such that each velocity 
 range reveals distinct features that appear to be similar (symmetrical) in both lobes of the molecular outflow.
 These features are clearly separated in velocity (and space) and do not appear be a single continuous structure in the outflow. 
At low velocities  the wide-angle bipolar outflow reported by previous studies is clearly seen. 
Our map reveals that the low-velocity blueshifted lobe has a parabolical morphology, 
while the redshifted lobe is a conical (or triangular) shape.  The vertices of the cones 
have a projected opening angle of about  90$^\circ$ for the blue lobe and 
about 80$^\circ$ for the red lobe. At intermediate velocities the outflow 
shows parabolic lobes that have  narrower opening angles than the 
low-velocity outflow gas; we measured an opening angle of 50$^\circ$ for the red and  60$^\circ$ for the blue lobe.
 At high velocities we detect very compact structures (or bullets) in the 
 blue and red lobes that are equidistant from the position of the source. 
Including the maps of all three different velocity ranges, for both lobes, together in the
 same figure  the outflow shows  a ``spider-like'' structure (see Figure 1b). 

In Figure 1b, we also included the positions of the strongest 
knots from the jet mapped in H$_2$ and H$\alpha$ by \citet{yu1999}. 
There is a good correspondence between the molecular and optical knots 
with a small offset of about 5$''$ (or $\sim 1200$ AU) between corresponding peaks,  with both
molecular bullets being farther away from the source than the optical/IR knots. 
One possible explanation to this offset is that the knots and the molecular 
bullets are produced by the same mass ejection episode, and that the difference 
in position is caused by the fact that the two different datasets were taken approximately 
a decade apart. A tangential velocity of approximately 570 km s$^{-1}$ would cause the 
observed positional offset between the optical knot and the 
molecular bullet.  Such very high velocities have been observed in several 
Herbig-Haro objects \citep[e.g. HH 111; HH 39 ][]{jon1982, sch1984, har2001}. 
We note that both the blueshifted and redshifted bullets are 
almost symmetrical in position (with respect to the source) which suggests 
that they were both ejected in the same mass ejection event.  
We tentatively  
suggest that the optical and molecular bullets are different manifestations of the same mass ejection episode. 
However, more optical/IR and
millimeter data are needed to confirm this.  

Assuming local thermodynamic equilibrium (LTE), and that the $^{12}$CO(2-1) molecular emission 
is optically thin, we estimate the outflow mass using the following equation:
$$
\frac{M(H_2)}{M_\odot}=6.3 \times 10^{-20} m(H_2) X_\frac{H_2}{CO}  \left (\frac{c^2  d^2}{2k\nu^2}  \right ) \frac{ \exp \left ( \frac{5.5}{T_{ex}} \right ) \int
S_{\nu} dv \Delta \Omega}{ \left (1 - \exp \left[\frac{-11.0}{T_{ex}} \right ] \right ) }, 
$$ 
where all units are in cgs, m(H$_2$) is mass of the molecular hydrogen with a value of 3.34$\times$10$^{-24} gr$, X $_\frac{H_2}{CO}$ 
is the fractional abundance between the carbon monoxide and the molecular hydrogen (10$^4$), $c$ is the speed of light with a value of 3$\times$10$^{10}$ cm s$^{-1}$, $k$ is the Boltzmann constant with a value of 1.38$\times$10$^{-16}$ 
erg K$^{-1}$, $\nu$ is the rest frequency of the CO line in Hz, $d$ is distance (240 pc), a parsec is equivalent to 3.08 $\times$10$^{18}$ cm, 
$S_{\nu}$ is the flux density of the CO (Jansky), a Jansky is equivalent to 1.0 $\times$10$^{-23}$ erg s$^{-1}$ cm$^{-2}$ Hz$^{-1}$,  $dv$ is the velocity range in cm s$^{-1}$, 
$ \Delta \Omega$ is the solid angle of the source in steradians,   
and T$_{ex}$ is excitation temperature taken to be 50 K.  Using only emission above 4-$\sigma$ in every velocity channel  map, we 
estimate a mass for the outflow powered  by B5-IRS 1 to be  7$\times$10$^{-3}$ M$_\odot$. 
This value is consistent with the mass of other molecular outflows powered 
by young low-mass protostars, see \citet{wu2004}. The mass estimated here is only a lower limit because the CO emission is likely to be optically thick
\citep[see, e.g.,][]{dun2014}. 
% We used a threshold of 4-$\sigma$ in every channel velocity map in order to estimate the mass.

We also estimate a kinematical energy ($E_k=\sum_{i} \frac{1}{2} \cdot m \cdot v_i^2$) of  2 $\times$ 10$^{44}$ ergs, 
and an outflow momentum ($p=\sum_{i}  m \cdot v_i$) of 0.5 M$_\odot$ km s$^{-1}$. From Figure 1, we estimate a dynamical age approximately of 700 years, which results 
in a  mechanical luminosity ($L=\frac{E_k}{t}$, where $E_k$ is the kinematical energy and $t$ is the dynamical age) of about 2 L$_\odot$.  
The dynamical age is estimated assuming a size for the molecular outflow of 30$''$ and a velocity of 50 km s$^{-1}$. 
These estimates do not take into consideration the inclination of the outflow with respect to the line-of-sight 
(which is thought to be high for this source, see Yu et al. 1999).

In Figure 3, we show the integrated intensity map of the SO(6$_5$-5$_4$) and the $^{13}$CO(2-1) emission. 
The SO line emission reveals a single compact
object associated with the continuum source in the position of B5-IRS 1, and does not show any clear velocity gradient in our data.
On the other hand, the $^{13}$CO(2-1) emission reveals cone-like lobes in the inner region of the bipolar outflow. 

\subsection[]{ Shells from a spider-like outflow}

In Figure 4  we present the Positional-Velocity Diagram (PVD) of the $^{12}$CO(2-1) emission 
along the outflow axis (P.A. = 67.1$^\circ$).
The PVD shows two structures at different velocity ranges. The main structure in the middle of the diagram,
at low to intermediate velocities (within  outflow velocities of 30 
km s$^{-1}$), shows a rotated ``{\it X}" morphology. 
Comparing the molecular outflow properties predicted by different models presented in \citet{arce2007} with
the velocity structure (from the PVD) and the cone-like morphology in both lobes of the B5-IRS 1 outflow (in Figure 1{\it a})
 we propose that this outflow is mostly driven by a wide-angle wind, as concluded by \citet[][]{lan1996}. 

There is also evidence, 
from the optical and infrared observations by \citet{yu1999} and our SMA data ({\it i.e.},  
the high-velocity bullets shown in Figure 1{\it b}), that B5-IRS 1 powers a very collimated (jet-like) wind.
 The molecular outflow jet component is also observed in our PVD as a structure at high outflow velocities 
 distinct from the central rotated ``{\it X}" structure in Figure 4. This is most evident in the redshifted lobe 
where faint emission, at offsets more negative than $-10\arcsec$ and at LSR velocities greater than $+$ 20 km s$^{-1}$
(and up to about  $+$ 60 km s$^{-1}$), shows outflow velocity increasing with offset from the source, 
as expected in a jet-driven molecular outflow \citep[see, {\it e.g.},][]{lee2000, lee2001}. 
The PVD in Figure 4 shows that the molecular bullets in both lobes reach outflow velocities of about  50 km s$^{-1}$.

In Figure 5 we present position-velocity diagrams of both outflow lobes, along cuts
 perpendicular to the outflow axis  (P.A. = 157.1$^\circ$). 
 These PVDs present different structures compared to those shown in Figure 4. 
 At low outflow velocities ($\lesssim 10$~km s$^{-1}$)
there are two bright curved structures that have similar velocity ranges in 
both outflow lobes, and are related to the low-velocity and very wide-angle structures seen in Figure 1{\it b}. 
In addition, there are two (faint) ring-like structures  in both sides of the flow at intermediate velocities (at 
outflow velocities of about 10 to 45~km s$^{-1}$). Such structures, in position-velocity space, 
are reminiscent of radially expanding shells or bubbles
\citep[see for example][]{zap2011, arce2011}, but are also similar to the elliptical structures 
expected in the PVD of a jet with a low inclination with respect to the plane of the sky
%\footnote {We note that \citet{yu1999} obtained a rough estimate for the inclination 
%angle between the outflow axis and the plane of the sky of about $13^\circ$,
%using the geometric model for the motions of CO outflow presented by \citep{liseau1986}}
\citep[see Figure 26 of][]{lee2000}. 
As discussed above, the integrated intensity map at   intermediate velocities show 
narrow parabolic structures, very different from  the circular (or semi-circular) 
structures associated with radially expanding bubbles. Furthermore, the observed structure 
is somewhat similar to the structures seen in 
in young (Class 0) jet-powered molecular outflows like L1448 \citep{bachi1995}, L1157 \citep{gueth1996} and 
HH 211 \citep{gueth1999,  pal2006, lee2007}. In these young sources the molecular outflow cavities have a moderate opening angle 
%(about 15$^\circ$ to 40$^\circ$) 
(similar to the intermediate-velocity structures we detect in B5-IRS1)
that are presumably formed by the passage of jet bow-shocks,  consistent with the 
  bow shock-driven molecular outflow model of \citet{raga1993}.
It therefore seems more likely that the observed 
intermediate-velocity features are caused by propagating bow shocks in a jet, or highly collimated wind. 
%radially expanding parabolic shell, 
%as predicted by the models shown by \citet{lee2000}. %As indicated by
%\citet{vel1998}, this source shows clear evidence that the outflow opening angle
%has increased with time, and therefore it is likely that the narrower intermediate-velocity feature was produced by 
%an earlier mass ejection episode (see below).
The high-velocity bullets
are  located at the tips of the intermediate-velocity elliptical structure in the  
position-velocity diagrams shown in Figure 5 (and at the far ends of the PVD shown in Figure 4).
Hence, it seems possible that these high-velocity features are tracing the tip (or head) of the bow-shock responsible for
the intermediate-velocity cavities.

\subsection[]{Nature of the morphology of the B5-IRS1 outflow}

One of the most striking characteristics of the  B5-IRS1 outflow is the different structures observed at 
different ranges of velocities. As discussed above, the high-velocity bullets and intermediate velocity structure are driven by a 
collimated (jet-like) wind,  while the low-velocity lobes are consistent with  being formed by
a wide-angle wind. 
An increasing number of  outflows have been observed that show two components  
(a wide-angle structure and a collimated feature). They include  L1551 \citep{ito2000}, 
HH 46/47 \citep{vel2007,arce2013}, 
%HH 211 \citep{lee2007, hir2006, pal2006}, 
Cepheus HW2 \citep{tor2011}, 
and Source I \citep{mat2010, zap2012}.  
However, to our knowledge, none of these outflows show a spider-like structure like the one we
observe in B5-IRS 1.

These ``dual component'' molecular outflows are generally explained by assuming that the underlying 
protostellar wind that entrains the surrounding ambient gas has a wide-angle morphology with a
narrow component (with a much higher outflow momentum rate) along the wind axis  
%wide component and a narrow 
%narrow component
%along the outflow axis and a wider, (and possibly slower moving or less dense), part
 as in the X-wind models of \citet{sha2006} and 
disc-magnetosphere boundary outflow launching models of \citet{rom2009}. 
The spider-like morphology of the B5-IRS 1 molecular outflow may be explained with 
these  kinds of models.  One hypothesis for the observed evolutionary trend in outflow 
opening angle is that the observed molecular outflow (produced by the interaction between 
the protostellar wind and the surrounding ambient gas) is mostly driven by one of the two 
different components, depending on the evolutionary stage of the protostar and the density 
distribution of the circumstellar material. During the early deeply embedded stage of protostellar evolution
(Class 0) only the high-momentum component along the outflow axis is able to puncture through 
the dense circumstellar envelope, producing a collimated (jet-like) outflow. 
At later stages ({\it i.e.}, Class I and II), after the envelope looses mass through outflow 
entrainment and infall onto the protostar, the wider component will be 
able to entrain the remaining circumstellar material at
larger angles away from the outflow axis. If such picture is correct,  
we should then expect a transitional phase when there is little or no molecular gas along the outflow axis, yet
enough molecular gas at intermediate angles  ({\it i.e.}, between the outflow axis and the edge of the wide-angle wind) 
that would result in a molecular outflow with a
component that is approximately midway 
between the collimated and wide-angle components, and a faint (or no) on-axis component.

We suggest this  scenario explains the spider-like structure
seen in B5-IRS1, which in fact may be considered a young
Class I source (based on its spectral energy distribution, e.g., Arce \& Sargent 2006) and hence
likely to be in an evolutionary phase close to the transition between Class 0 and Class I (see also Yu et al.~1999).
We argue that in B5-IRS1 the molecular outflow material  is detected along (and within 30$^\circ$~of) the outflow axis (as shown by the high-velocity bullet and the intermediate-velocity structure) because there is still enough molecular material in the cavity for it to be entrained by the collimated (jet) component of the wind
(as expected in a young source). However,  as seen in Figure 1 (where the integrated intensity of the outflowing CO emission is dominated by the very wide angle structure) most of the entrainment is taking place at larger angles from the axis where the wide-angle wind is currently interacting with the 
denser parts of the surrounding envelope. 

%In this scenario, for a source in such a transitional phase and with a constant mass outflow rate, one would expect a smooth change in morphology with 
%decreasing velocity ---from a collimated outflow to a wide-angle structure. However, in the case of an outflow with episodes of very high mass loss rates between periods of low outflow activity, the  velocity-dependent morphology would not necessarily show such smooth transition.
%In B5-IRS 1 the distribution of HH objects and H$_2$ knots along the outflow axis, as well as the morphology and kinematics of the  outflowing gas 
%indicate that this is an episodic outflow \citep{yu1999}. 
%In addition, B5-IRS 1 is thought to be a young Class I source (based on its spectral energy distribution, e.g., Arce \& Sargent 2006) 
%and hence likely to be in an evolutionary phase close to the transition between Class 0 and Class I (see also Yu et al.~1999). 
%{\bf One therefore would think that the spider-like structure in B5-IRS 1 is the result of an episodic outflow, however, this suggestion seems to be incompatible
%because of the dynamical times for the bullets and wide opening angles observed within the flow.} 

\section*{Summary}

We  observed in the millimeter regime the dust and molecular gas surrounding the young stellar object B5-IRS 1 
using the Submillimeter Array.  Our conclusions are as follow:

\begin{itemize} 

\item The millimeter dust continuum emission reported here is tracing the  
          envelope and circumstellar disk surrounding B5-IRS 1.

\item Our $^{12}$CO(2-1) observations resolve the bipolar northeast-southwest outflow  
         associated with B5-IRS 1 and find that its morphology is reminiscent of a ``spider", where 
          three velocity  components with different morphologies are present in each lobe.   
           
\item  In addition to detecting the previously observed wide-angle cone-like lobes, our observations  
        for the first time reveal 
        the presence of intermediate-velocity,  parabolic shells emerging very close 
         to the young stellar object as well as high-velocity compact molecular bullets  
         which we argue are associated with the optical/IR jet in this source.
          These high-velocity features reach outflow (radial) velocities of about 50 km s$^{-1}$.

\item  We interpret the peculiar spider-like morphology as a result of the molecular 
          material being entrained by a wind in which the momentum 
          has an angular dependence (i.e., larger towards the outflow axis).  We believe
          the peculiar outflow morphology is evident in this source because  it is in a transitional evolutionary
          phase, at a stage that is slightly older than the phase in which the outflow is completely dominated 
          by the on-axis (collimated) part of the wind, but slightly younger than the stage in which 
          the outflow entrainment is fully dominated by the wide-angle component of the wind.

\item  We report the detection $^{13}$CO(2-1) and SO(6$_5$-5$_4$) emission, 
	 which arises from the outflow and the vicinity of B5-IRS1.
          
\end{itemize} 

We conclude that the protostar B5-IRS 1 is a great laboratory to study the process of 
outflow formation and evolution. Further observations at different wavelengths and at higher 
angular and velocity resolution are needed to better understand its kinematics as well its launching 
and entrainment mechanisms. 

\section*{Acknowledgments}

L.A.Z. acknowledge the financial support from DGAPA, UNAM, and CONACyT, M\'exico.
H.G.A. acknowledges support from his NSF CAREER award AST-0845619. A.P. is supported 
by the Spanish MICINN grant AYA2011-30228-C03-02 (co-funded with FEDER funds), 
and by the AGAUR grant 2009SGR1172 (Catalonia).

\end{document}